\pdfoutput=1

\documentclass[aps,prd,preprint,tightenlines,superscriptaddress,
amsfonts,amssymb,amsmath,byrevtex,showpacs,nofootinbib]{revtex4-1}

\usepackage{amssymb,amsmath}
\usepackage{latexsym}
\usepackage{graphicx}
\usepackage{geometry}
\usepackage{epstopdf}
\usepackage{setspace}
\usepackage{hyperref}
\usepackage{amsthm}
\usepackage{amsmath}
\usepackage{amssymb}
\usepackage{pgfplots}
\usepackage{graphicx} 
\usepackage{mathrsfs}
\usepackage{subcaption}
\usepackage{color}
\usepackage{verbatim}
\usepackage{tensor}
\usepackage{xcolor}
\usepackage{braket}

\newcommand{\dd}{\textnormal{d}}

\newcommand{\ee}{\textnormal{e}}

\begin{document}
\title{On the dynamics of the general Bianchi IX spacetime \\ near the
  singularity }

\author{Claus Kiefer}
\email{kiefer@thp.uni-koeln.de}\affiliation{Institute for Theoretical Physics,
University of Cologne, Z\"{u}lpicher Strasse 77, 50937 K\"{o}ln, Germany}

\author{Nick Kwidzinski}
\email{nk@thp.uni-koeln.de}\affiliation{Institute for Theoretical Physics,
University of Cologne, Z\"{u}lpicher Strasse 77, 50937 K\"{o}ln, Germany}

\author{W{\l}odzimierz Piechocki}
\email{wlodzimierz.piechocki@ncbj.gov.pl}\affiliation{ Department
of Fundamental Research, National Centre for Nuclear Research,
Ho{\.z}a 69, 00-681 Warszawa, Poland}

\date{\today}
\begin{abstract}
We show that the complex dynamics of the general Bianchi IX universe in the vicinity
of the spacelike singularity can be approximated by a simplified
system of equations.
Our analysis is mainly based on numerical simulations. The
  properties of the solution space can be studied by using this simplified dynamics.
Our results will be useful for the quantization of the general Bianchi IX
model.
\end{abstract}


\maketitle

\section{Introduction}

The problem of spacetime singularities is a central one in classical
and quantum theories of gravity. Given some general conditions, it was proven
that general relativity leads to singularities, among which special
significance is attributed to big bang and black hole singularities
\cite{HP96}.

The occurrence of a singularity in a physical theory usually signals
the breakdown of that theory. In the case of general relativity, the
expectation is that its singularities will disappear after
quantization. Although a theory of quantum gravity is not yet available in
finite form, various approaches exist within which the question of
singularity avoidance can be addressed \cite{oup}. Quantum cosmological
examples for such an avoidance can be found, for example, in
\cite{ABKMM,KKK16,Bergeron:2015lka,Bergeron:2015ppa} and the references therein.

Independent of the quantum fate of singularities, the question of
their exact nature in the classical theory, and in particular for
cosmology, is of considerable interest
and has a long history; see, for example, \cite{Berger} and
\cite{Uggla} for recent reviews. This is also the topic of the present
paper.

Already in the 1940s, Evgeny Lifshitz investigated the gravitational stability of
non-stationary isotropic models of universes.
He found that the isotropy of space cannot be retained  in the
evolution towards singularities \cite{Lif1} (see \cite{Lif2}
for an extended physical interpretation). This motivated the activity
of the Landau Institute in Moscow to examining
the dynamics of homogeneous spacetimes \cite{Bel}. A group of
relativists inspired by Lev Landau, including Belinski,
Khalatnikov and Lifshitz (BKL), started to investigating
the dynamics of the Bianchi VIII and IX models near
the initial spacelike cosmological singularity
\cite{BKL1}. After several years, they found that the dynamical behaviour
 can be generalized to a generic solution of general
relativity \cite{BKL2}. They did not present a mathematically
rigorous proof, but rather a conjecture based on deep analytical
insight. It is called the BKL conjecture
(or the BKL scenario if specialized to the Bianchi-type IX model).
The BKL conjecture is a locality conjecture stating
that terms with temporal derivatives dominate over terms with spatial
derivatives when approaching the singularity
(with the exception of possible `spikes'
\cite{Uggla,Czuchry:2016rlo}). Consequently, 
points in space decouple and the dynamics then
turn out to be effectively the same as those of the (non-diagonal)
Bianchi IX universe. (In canonical gravity, this is referred to as the
strong coupling limit, see e.g. \cite{oup}, p.~127.)

The dynamics of the Bianchi
IX towards the singularity are characterized by an infinite number of
oscillations, which give rise
to a chaotic character of the solutions (see e.g
\cite{Cornish:1996hx}). Progress towards improving the mathematical
rigour of the BKL conjecture has been made by several authors (see e.g.
\cite{Heinzle_Uggla_Rohr_2009}), while
numerical studies giving support to the conjecture have
been performed  (see e.g. \cite{Garfinkle_2004}).

The dynamics of the diagonal Bianchi IX model, in the Hamiltonian
formulation, were studied independently from BKL by Misner
\cite{Misner_1969a,Misner_1969b}. Misner's intention was to
search for a possible solution to the horizon problem
by a process that he called ``mixing''.\footnote{This is how the
  diagonal Bianchi IX model received the name mixmaster
  universe.}
Ryan generalized Misner's formalism to the non-diagonal case in
\cite{Ryan_1971a,Ryan_1971b}.
A qualitative treatment of the dynamics for all the Bianchi
models may be found in the review article by Jantzen
\cite{Jantzen:2001me}, and we make reference to it whenever we get
similar results.

Part of the BKL conjecture is that non-gravitational (`matter')
terms can be neglected when approaching the singularity. An important
exception is the case of a massless scalar field, which has analogies
with a stiff fluid (equation of state $p=\rho$) and, in Friedmann models, has the same
dependence of the density on the scale factor as anisotropies ($\rho\propto a^{-6}$). As
was rigorously
shown in \cite{AR01}, such a scalar field will suppress the
BKL oscillations and thus {\em is} relevant during the evolution
towards the singularity. Arguments for the importance of stiff matter
in the early universe were already given by Barrow \cite{Barrow78}.

In our present work, we shall mainly address the general
(non-diagonal) Bianchi~IX model near its singularity.
Our main motivation is to provide support for a rather simple
asymptotic form of the dynamics that can suitably model
its exact complex dynamics. We expect this to be of relevance in the
quantization of the general Bianchi~IX model, which we plan to investigate in later
papers; see, for example, \cite{AGWP}.
Apart from a few particular solutions that form a set of measure zero in
the solution space, no general analytic
solutions to the classical equations of motion are known. Therefore,
we will restrict ourselves to qualitative
considerations which will be supported by numerical simulations.
The examination of the non-diagonal dynamics presented in \cite{Bogo2},
though it is mathematically satisfactory, is based on the
qualitative theory of differential equations, which is of little
use for our purpose.

Our paper is organized as follows.
Section II contains the formalism and presents our main results for a general
Bianchi~IX model.
We first specify the kinematics and dynamics. We then consider a matter
field in the form of (tilted) dust. This is followed by
investigating the asymptotic regime
of the dynamics near the singularity.
Our conclusions are presented in Sec. III. The numerical methods used
in our numerical
simulations are described in the Appendix.

\section{The general Bianchi IX spacetime}

\subsection{Kinematics}

The general non-diagonal case  describes a universe with rotating
principal axes.
The metric in a synchronous frame can be given as follows
(see for the following e.g. \cite{Ryan_Shepley} and
\cite{Ryan_HC}):
\begin{equation}
\dd s^2=-N^2 \dd t^2 + h_{ij} \sigma^i \otimes\sigma^j ,
\label{eq:Non_diagonal_BIX_metric}
\end{equation}
where $N$ is the lapse function.
Spatial hypersurfaces in the spacetime are regarded topologically as
$S^3$ (describing closed universes),
 which can be parametrized by using three angles
$\left\{\bar{\theta},\bar{\phi},\bar{\psi} \right\}\in  [0,\pi]\times[0,2\pi]\times [0,\pi]$. The basis one-forms read
\begin{equation}
\begin{aligned}
\sigma^1 &=  -\sin (\bar{\psi}) \dd \bar{\theta} + \cos (\bar{\psi})\sin(\bar{\theta}) \dd \bar{\phi}\, ,
\\
\sigma^2 &=  \cos (\bar{\psi})  \dd \bar{\theta} + \sin (\bar{\psi})\sin(\bar{\theta}) \dd \bar{\phi}\, ,
\\
\sigma^3 &=  \cos(\bar{\theta})\dd \bar{\phi} +  \dd \bar{\psi}
\ .
\end{aligned}
\end{equation}
The $\sigma^i$, $i=1,2,3$, are dual to the vector fields
\begin{equation}
\begin{aligned}
X_1 &=  -&\sin(\bar{\psi}) \partial_{\bar{\theta}} +
\frac{\cos(\bar{\psi})}{\sin (\bar{\theta})}
\left[
\partial_{\bar{\phi}}
+ \cos(\bar{\theta})  \partial_{\bar{\psi}}
\right]\, ,
\\
X_2 &= &\cos(\bar{\psi}) \partial_{\bar{\theta}} +
\frac{\sin(\bar{\psi})}{\sin (\bar{\theta})}
\left[
\partial_{\bar{\phi}}
+ \cos(\bar{\psi})  \partial_{\bar{\psi}} \right]\, ,
\\
X_3 &=& \partial_{\bar{\psi}}\, ,
\\
\end{aligned}
\end{equation}
which together with $X_0=\partial_t$ form an invariant basis of the
Bianchi IX spacetime. The $X_i$ are
constructed from the Killing vectors that generate the isometry group
$SO(3,\mathbb{R})$ (see \cite{Ryan_Shepley} for more details).
The basis one-forms satisfy the relation
\begin{equation}
\dd \sigma^i=-\frac{1}{2}C^i_{jk}\sigma^j \wedge \sigma^k \ ,
\end{equation}
with $C^i_{jk}=\varepsilon_{ijk}$ being the structure constants of the Lie algebra $\mathfrak{so}(3, \mathbb{R})$. The $X_i$
obey the algebra $[X_i,X_j]=-C^{k}_{ij}X_k$.
We parametrize the metric coefficients  in this frame as follows
\begin{equation}
h_{ij} =  O_i{}^{k}  O_j{}^{l} \bar{h}_{ k l }\, ,
\end{equation}
where
\begin{equation}
\label{bar-h}
\bar{h}\equiv\left\{
\bar{h}_{i j}
\right\} =
 \ee^{2\alpha}\text{diag}\left(
\ee^{2\beta_+ + 2\sqrt{3}\beta_-},
\ee^{2\beta_+ - 2\sqrt{3}\beta_-},
\ee^{-4\beta_+ }
\right)\equiv\text{diag}\left(\Gamma_1 ,\Gamma_2 , \Gamma_3 \right) .
\end{equation}
The variables $\alpha$, $\beta_+$, and $\beta_-$ are known as the
Misner variables. The scale factor $\exp(\alpha)$ is related to the volume,
while the anisotropy factors $\beta_+$ and $\beta_-$ describe the shape
of this model universe.
The variables $\Gamma_1$, $\Gamma_2$  and  $\Gamma_3 $ were used by
BKL in their original analysis \cite{bkl}.

We introduced here a matrix $O\equiv\left\{O_i{}^{j}\right\}\equiv
O_{\theta}O_{\phi}O_{\psi}$
($i$ corresponding to rows and $j$ corresponding to columns), which is an
$SO(3,\mathbb{R})$ matrix
that can be parametrized by another set of Euler angles,
$\left\{ \theta,\phi,\psi \right\}\in
[0,\pi]\times[0,2\pi]\times [0,\pi]$. Explicitly,
\begin{equation}
\label{matrixO}
\begin{aligned}
O_{\psi}=\left(
\begin{array}{ccc}
\cos ( \psi ) & \sin ( \psi ) & 0 \\
- \sin ( \psi ) & \cos ( \psi ) & 0 \\
0 & 0 & 1
\end{array}
\right) \ , \quad
O_{\theta}=\left(
\begin{array}{ccc}
1 & 0 & 0 \\
0 &\cos ( \theta ) & \sin ( \theta )  \\
0 &- \sin ( \theta ) & \cos ( \theta )
\end{array}
\right)\, ,
\\
O_{\phi}=\left(
\begin{array}{ccc}
\cos ( \phi ) & \sin ( \phi ) & 0 \\
- \sin ( \phi ) & \cos ( \phi ) & 0  \\
0 & 0 & 1
\end{array}
\right)\, .\qquad\qquad\quad\qquad
\end{aligned}
\end{equation}
The Euler angles $\theta$, $\phi$, and $\psi$ are now
dynamical quantities and describe nutation,
precession, and pure rotation of the principal axes, respectively. In
  the case of Bianchi IX spacetime, the group $SO(3,\mathbb{R})$ is the canonical
  choice for the diagonalization of the metric coefficients. For a
  treatment of other Bianchi models, see \cite{Jantzen:2001me}.

\subsection{Dynamics}

 In the following, we shall discuss the Hamiltonian formulation of this model.
In order to keep track of the diffeomorphism (momentum) constraints, we replace the
metric (\ref{eq:Non_diagonal_BIX_metric}) by the ansatz
\begin{equation}
\begin{aligned}
&\dd s^2=- N^2 \dd t^2 +
   h_{ij}\left(N^i\dd t+\sigma^i \right)\otimes\left(N^j\dd t+\sigma^j \right),
\end{aligned}
\end{equation}
where $N^i$ are the shift functions.
The Hamiltonian formulation was first derived in a series of papers by
Ryan: the symmetric (non-tumbling) case obtained by constraining
${\psi}$, ${\phi}$ to be constant and keeping $\theta $ dynamical is
discussed in \cite{Ryan_1971a}, and the general case can be found in
\cite{Ryan_1971b}.
We write the Einstein-Hilbert action in the well known ADM form,
\begin{equation}
S_{EH}=
\frac{1}{16\pi G}\int \sigma^1\wedge\sigma^2\wedge \sigma^3\int\dd t\
N\sqrt{h}\left[
\left(h^{ik} h^{jl}  -h^{ij}h^{kl}\right)  K_{ij}K_{kl}
+{}^{(3)} R
\right],
\label{eq:SEH_BIX}
\end{equation}
where
\begin{displaymath}
K_{ij}=\frac{1}{2N}\left(\dot{h}_{ij}-2 D_{(i} N_{j}) \right)
\end{displaymath}
is the extrinsic curvature, and $D_i$ is the spatial covariant
derivative in the non-coordinate basis $\left\{ X_i \right\}$. We will
set $\frac{3}{4\pi G}\int \sigma^1\wedge\sigma^2\wedge \sigma^3=1$ for
simplicity.
The three-dimensional curvature  ${}^{(3)}R$ on spatial hypersurfaces of constant
coordinate time is given by
\begin{equation}
{}^{(3)} R =  -\frac{\ee^{-2\alpha}}{2}
 \left(
\ee^{-8\beta_+}-4\ee^{-2\beta_+} \cosh \left(2\sqrt{3}\beta_- \right)
+ 2\ee^{4\beta_+}
\left[ \cosh\left(
4\sqrt{3}\beta_-
\right)
-1
\right]
\right) .
\label{eq:3R_BIX}
\end{equation}
We now turn to  the calculation of the kinetic term and the
diffeomorphism constraints. For this purpose, we define an
antisymmetric angular velocity tensor $\omega^{i}{}_{j}$ by the
matrix equation
\begin{equation}
\boldsymbol{\omega}=\left\{\omega^{i}{}_{j} \right\}=\left(\begin{array}{ccc}
0 & \omega^{1}{}_{2} & -\omega^{3}{}_{1} \\
-\omega^{1}{}_{2} & 0 & \omega^{2}{}_{3} \\
\omega^{3}{}_{1} & -\omega^{2}{}_{3} & 0
\end{array}  \right)   \equiv O^{T}\dot{O}.
\end{equation}
An explicit calculation of the right-hand side gives, using \eqref{matrixO},
\begin{align}
\omega^{2}{}_{3} = & \cos(\psi) \dot{\phi} +\sin(\psi)\sin(\phi)\dot{\theta}  \ ,
\\
\omega^{3}{}_{1} = & \sin (\psi)\dot{\phi}-\cos(\psi)\sin(\phi)\dot{\theta}  \ ,
\\
\omega^{1}{}_{2} = & \dot{\psi}+\cos(\phi)\dot{\theta}  \ .
\end{align}
The Lagrangian in the gauge $N^i=0$  then takes the form
\begin{equation}
\text{L}=N \ee^{3\alpha}\left[\frac{-\dot{\alpha}^2+\dot{\beta}^2_+ + \dot{\beta}^2_-
+   I_1 \left(\omega^{2}{}_{3}\right)^2
+   I_2 \left(\omega^{3}{}_{1}\right)^2
+   I_3 \left(\omega^{1}{}_{2}\right)^2
}{2N^2}
+ \frac{{}^{(3)} R}{12}\right] ,
\end{equation}
where the `moments of inertia' are given by
\begin{equation}
3 I_1  \equiv\sinh^2 \left( 3\beta_+ - \sqrt{3}\beta_-  \right)  , \quad
3 I_2  \equiv \sinh^2 \left( 3\beta_+ + \sqrt{3}\beta_-  \right)  , \quad
3 I_3  \equiv \sinh^2 \left( 2\sqrt{3}\beta_- \right) .
\end{equation}
Note, in particular, that the term
$\frac{1}{2}\left[ I_1 \left(\omega^{2}{}_{3}\right)^2
+   I_2 \left(\omega^{3}{}_{1}\right)^2
+   I_3 \left(\omega^{1}{}_{2}\right)^2  \right]$
would formally correspond to the rotational energy of a rigid body if the
moments of inertia were constant.
The canonical momenta conjugate to the Euler angles are given by
\begin{equation}
\begin{aligned}
p_{\theta}
& =
\frac{\ee^{3\alpha}}{N}\left[
I_1 \sin(\psi )\sin (\phi)\omega^{2}{}_{3}-I_2\cos(\psi )\sin (\phi)\omega^{3}{}_{1}
+I_3 \cos(\theta)\omega^{1}{}_{2}
\right],
\\
p_{\phi}
& =
\frac{\ee^{3\alpha}}{N}\left[
I_1 \cos (\psi)\omega^{2}{}_{3}
+I_2 \sin (\psi)\omega^{3}{}_{1}
\right],
\\
p_{\psi}
& =
\frac{\ee^{3\alpha}}{N}\ I_3 \omega^{1}{}_{2}.
\end{aligned}
\end{equation}
It is convenient to introduce the following (non-canonical) angular momentum
like variables:
\begin{equation}
 l_1 \equiv\frac{\ee^{3\alpha}}{N}I_1 \omega^{2}{}_3  , \quad
 l_2 \equiv\frac{\ee^{3\alpha}}{N}I_2 \omega^{1}{}_3    \quad \text{and} \quad
 l_3 \equiv\frac{\ee^{3\alpha}}{N}I_3 \omega^{2}{}_1 .
\end{equation}
The relation to the canonical momenta can now explicitely be given by
\begin{equation}
\begin{gathered}
p_{\theta}
=
\sin (\psi)\sin(\phi) l_1 - \cos(\psi)\sin(\phi) l_2 +\cos(\phi) l_3 \ ,
\\
p_{\phi}
=
\cos(\psi) l_1 +\sin(\psi) l_2 \ ,
\\
p_{\psi}
=
l_3 \ .
\end{gathered}
\end{equation}
It is readily shown that the variables $l_i$ obey the Poisson bracket algebra
$
\left\{ l_i,l_j \right\} =-C^k_{ij}{}l_k
$.
After the usual Legendre transform, we obtain the Hamiltonian constraint,
\begin{equation}
\mathcal{H} =
\frac{\ee^{-3\alpha}}{2}\left(-p_\alpha^2 +p_+^2 +p_-^2
  +\frac{l_1^2}{I_1}+\frac{l_2^2}{I_2}+\frac{l_3^2}{I_3}  -
  \frac{\ee^{6\alpha}}{6} {}^{(3)}R \right)  .
\end{equation}
From  (\ref{eq:SEH_BIX}), we find that the diffeomorphism constraints
($\partial L/\partial N^i=0$)  can be written as
\begin{equation}
\mathcal{H}_i = 2C^{j}{}_{il} h_{jk}p^{kl} ,
\end{equation}
where
\begin{displaymath}
p^{ij}=\frac{\sqrt{h}}{24N}\left( h^{ik}h^{jl}-h^{ij}h^{kl}
\right)K_{kl}
\end{displaymath}
 is the ADM momentum.
From this expression we can finally compute the diffeomorphism constraints in terms of the angular momentum-like variables and obtain
\begin{equation}
\label{momentum-constraints}
\mathcal{H}_i=O_{i}{}^{j}l_j,
\end{equation}
that is, we can identify the diffeomorphism constraints with a basis
of the generators of $SO(3,\mathbb{R})$.
The full gravitational Hamiltonian then reads
\begin{equation}
H=N\mathcal{H}+N^i\mathcal{H}_i.
\end{equation}
From the diffeomorphism constraints \eqref{momentum-constraints} we
conclude that in the vacuum case $l_i=0$ and that therefore no rotation is possible,
that is, we recover
the diagonal case. If we want to obtain a Bianchi IX universe with
rotating principal axes, we are thus forced to add matter to
the system. A formalism
for obtaining equations of motion for general Bianchi class A models
filled with fluid matter was developed by Ryan \cite{Ryan_HC}.
For simplicity, we will only consider the case of dust as discussed
by Kucha\v{r} and Brown in
\cite{Kuchar_Brown_1995}.
 If we were, for example, interested in the study of the quantum
version of this model, it would be
desirable to introduce a fundamental matter field instead of an ideal fluid.
Standard scalar fields alone cannot lead to a rotation for Bianchi IX
models. The easiest
way to achieve this is, to our knowledge, the introduction of a Dirac field
 \cite{Damour_Spindel_2011}.

\subsection{Adding dust to the system}
The energy momentum tensor for dust reads $T_{\mu\nu} = \rho u_\mu u_\nu$.
The local energy conservation  $\nabla_\mu T^{\mu\nu}=0$ leads to a
geodesic equation for the positions of the dust particles.
Let us start therefore by considering the geodesic equation for
a single dust particle, whose four-velocity we can express in the
non-coordinate frame $\sigma^i$ used above by the Pfaffian form
\begin{equation}
\mathbf{u}=u_0 \dd t + u_i \sigma^i
\quad \text{with} \ \
\langle \mathbf{u},\mathbf{u} \rangle=-1\ .
\label{eq:BIX_dust_form1}
\end{equation}
We  partially fix the gauge by setting $N^i=0$.
The normalization condition implies
\begin{equation}
u_0=-  N \sqrt{1+h^{ij}u_i u_j}.
\label{eq:u_0}
\end{equation}
We have chosen here the minus sign because this guarantees that the proper
time in the frame of the dust particle has the same orientation as the
coordinate time $t$.
The geodesic equation for the spatial components of the four velocity
can then be written as
\begin{equation}
u^0(\partial_t u_i) - C^k_{ij}u_k u^j = 0 \ .
\label{eq:u_geodesic}
\end{equation}
The geodesic equation implies the existence of a constant of
motion. To see this explicitly, we compute the expression
$\sum\limits_{i=1,2,3}\dot{u}_i u_i$ and convince ourselves that it vanishes identically.
Thus the Euclidean sum
\begin{equation}
C^2\equiv(u_1)^2+(u_2)^2+(u_3)^2
\label{eq:u_geodesic_constant}
\end{equation}
is a constant of motion. Defining $\vec{u}\equiv(u_1, \ u_2, \ u_3)^{T}$, the
geodesic equation (\ref{eq:u_geodesic}) can be rewritten in
vector notation,
\begin{equation}
\partial_t \vec{u}= \frac{N \left[ \vec{u}\times (O \bar{h}^{-1} O^T\vec{u})\right] }{\sqrt{1+\vec{u}^T O \bar{h}^{-1} O^T \vec{u}}} \ ,
\end{equation}
where ``$\times$'' denotes the usual cross product in the three
dimensional Euclidean space.
Defining for convenience $\vec{v}\equiv O^T \vec{u}/C$, we have
$(v_1)^2+(v_2)^2+(v_3)^2 = 1$, and the
geodesic equation simplifies to
\begin{equation}
\left( \partial_t+\boldsymbol{\omega} \right)\vec{v} \ =\frac{ N C  \left[ \vec{v}\times ({{\bar{h}}}^{-1}\vec{v} )\right]}{\sqrt{1+C^2  \vec{v}^T {{\bar{h}}}^{-1}\vec{v} }}
\label{eq:v_geodesic} .
\end{equation}
Note that we can also write
$\boldsymbol{\omega}\vec{v}=\vec{v}\times\vec{\omega}$, where
\begin{displaymath}
 \vec{\omega}\equiv\{ \omega^i \}= \frac{N}{\ee^{3\alpha}} \left(
   \frac{l_1}{I_1},\ \frac{l_2}{I_2} ,\ \frac{l_3}{I_3} \right)^T.
\end{displaymath}
 It will thus be possible to eliminate $\boldsymbol{\omega}$ from the
 geodesic equation by using the diffeomorphism constraints.

We now add homogeneous dust to the system.
The formalism developed in
\cite{Kuchar_Brown_1995} leads to the following form of the
Hamiltonian and diffeomorphism constraints for dust in a Bianchi~IX
universe,
\begin{equation}
\begin{aligned}
&\mathcal{H} + \mathcal{H}^{(m)}
=
\frac{\ee^{-3\alpha}}{2}\left(-p_\alpha^2 +p_+^2 +p_-^2
+\frac{l_1^2}{I_1}+\frac{l_2^2}{I_2}+\frac{l_3^2}{I_3}
- \frac{\ee^{6\alpha}}{6} {}^{(3)}R
+ 2\ee^{3\alpha} p_T \sqrt{1+h^{ij}u_i u_j}
\right),
\\
&\mathcal{H}_i  +\mathcal{H}^{(m)}_i
= O_i{}^j\left( l_j - C p_T v_j \right) ,
\end{aligned}
\label{eq:Hamiltonian+diffeo_constraints_dust}
\end{equation}
where $p_T$ denotes the momentum canonically conjugate to
$T$, where $T$ is the global `dust time'.
Since the Hamiltonian does not explicitly depend on $T$, the momentum
$p_T$ is a constant of motion.
The fact that $l_1^2+l_2^2+l_3^2$
 commutes with $\mathcal{H}$ implies that $l_1^2+l_2^2+l_3^2=(Cp_T)^2$
 is a conserved quantity. This is consistent with
\eqref{eq:u_geodesic_constant}. We note that a similar form of the constraints
\eqref{eq:Hamiltonian+diffeo_constraints_dust} was already presented in
\cite{Ryan_1971a,Ryan_1971b}. The formalism is not entirely canonical
and must be complemented by the geodesic equation (\ref{eq:u_geodesic}).

For our numerical purposes, it will be convenient to
rewrite the equations of
motion in the variables $\Gamma_i$ introduced in \eqref{bar-h}. We find
that the choice of the
variables $\log \Gamma_i$ allow for a better control over the error in
the Hamiltonian constraint. Moreover, we pick the quasi-Gaussian gauge
$N=\ee^{3\alpha}=\sqrt{\Gamma_1 \Gamma_2 \Gamma_3}$, $N^i=0$. Recall
that the singularity is reached in a finite amount of comoving time
(corresponding to the gauge $N=1$). The choice $N=\ee^{3\alpha}$
allows to resolve the oscillations in the approach towards the
singularity.
With these choices the Hamiltonian constraint becomes
\begin{equation}
\begin{aligned}
&-(\log\Gamma_1)^{\cdot}(\log\Gamma_2)^{\cdot}-(\log\Gamma_2)^{\cdot}(\log\Gamma_3)^{\cdot}-(\log\Gamma_1)^{\cdot}(\log\Gamma_3)^{\cdot}
\\
&+\Gamma_1^2+\Gamma_2^2+\Gamma_3^2-2(\Gamma_1\Gamma_2+\Gamma_3\Gamma_1+\Gamma_2\Gamma_3)
\\
&+ 24 \left[\frac{l_1^2}{I_1}+\frac{l_2^2}{I_2}+\frac{l_3^2}{I_3}
+ 2 |p_T|\sqrt{\Gamma_1\Gamma_2\Gamma_3+ C^2 \left(\Gamma_2\Gamma_3 v_1^2+\Gamma_1\Gamma_3v_2^2+\Gamma_1\Gamma_2 v_3^2\right) }   \right] = 0,
\end{aligned}
\end{equation}
where the moments of inertia are
\begin{equation}
I_1 = \frac{(\Gamma_3-\Gamma_2)^2}{12\Gamma_3\Gamma_2}  ,
\quad
I_2 = \frac{(\Gamma_1-\Gamma_3)^2}{12\Gamma_1\Gamma_3}  ,
\quad
I_3 = \frac{(\Gamma_1-\Gamma_2)^2}{12\Gamma_1\Gamma_2}  .
\end{equation}
The diffeomorphism constraints read
$l_i=p_T C v_i  $ and can be used to eliminate the angular momentum variables from the equations of motion.
These equations can then be written as
\begin{equation}
\begin{aligned}
(\log \Gamma_1)^{\cdot \cdot}
=&
(\Gamma_2-\Gamma_3)^2-\Gamma_1^2 +
2p_T'^2C^2\left[
\frac{\Gamma_1\Gamma_3(\Gamma_1+\Gamma_3)v_2^2}{(\Gamma_1-\Gamma_3)^3}
+\frac{\Gamma_1\Gamma_2(\Gamma_1+\Gamma_2)v_3^2}{(\Gamma_1-\Gamma_2)^3} \right]
\\
& +
\frac{p_T'(\Gamma_1 \Gamma_2 \Gamma_3+2 C^2 v_1^2 \Gamma_2\Gamma_3)}
{\sqrt{\Gamma_1\Gamma_2\Gamma_3+ C^2 \left(\Gamma_2\Gamma_3 v_1^2+\Gamma_1\Gamma_3v_2^2+\Gamma_1\Gamma_2 v_3^2\right) }},
\\
(\log \Gamma_2)^{\cdot \cdot}
= &
(\Gamma_3-\Gamma_1)^2-\Gamma_2^2 +
2p_T'^2C^2\left[
\frac{\Gamma_1\Gamma_2(\Gamma_1+\Gamma_2)v_3^2}{(\Gamma_2-\Gamma_1)^3}
+\frac{\Gamma_2\Gamma_3(\Gamma_2+\Gamma_3)v_1^2}{(\Gamma_2-\Gamma_3)^3} \right]
\\
& +
\frac{p_T'(\Gamma_1 \Gamma_2 \Gamma_3+2 C^2 v_2^2 \Gamma_1\Gamma_3)}
{\sqrt{\Gamma_1\Gamma_2\Gamma_3+ C^2 \left(\Gamma_2\Gamma_3 v_1^2+\Gamma_1\Gamma_3v_2^2+\Gamma_1\Gamma_2 v_3^2\right) }},
\\
(\log \Gamma_3)^{\cdot \cdot}
=&
(\Gamma_1-\Gamma_2)^2-\Gamma_3^2 +
2p_T'^2C^2\left[
\frac{\Gamma_1\Gamma_3(\Gamma_1+\Gamma_3)v_2^2}{(\Gamma_3-\Gamma_1)^3}
+\frac{\Gamma_3\Gamma_2(\Gamma_3+\Gamma_2)v_1^2}{(\Gamma_3-\Gamma_2)^3} \right]
\\
& +
\frac{p_T'(\Gamma_1 \Gamma_2 \Gamma_3+2 C^2 v_3^2 \Gamma_1\Gamma_2)}
{\sqrt{\Gamma_1\Gamma_2\Gamma_3+ C^2 \left(\Gamma_2\Gamma_3 v_1^2+\Gamma_1\Gamma_3v_2^2+\Gamma_1\Gamma_2 v_3^2\right) }},
\end{aligned}
\label{eq:BIX_nondiag_eoms_rot_dust}
\end{equation}
where we have set $p_T'\equiv12p_T$ for convenience. Note that these
equations are exact. (In \cite{BKL1}, the matter terms were neglected.)
We use the diffeomorphism  constraints to eliminate  $\vec{\omega}$ from the geodesic equation
(\ref{eq:v_geodesic}).
If expressed in the gauge $N=\ee^{3\alpha}$ and using the
$\Gamma_i$,  the geodesic equation \eqref{eq:v_geodesic} can be written as
\begin{equation}
\begin{aligned}
\dot{\vec{v}}= & C\vec{v}\times \left( M\vec{v} \right) \quad \text{where}
\\
M = &\frac{\text{diag}\left(\Gamma_2 \Gamma_3, \ \Gamma_1 \Gamma_3,\ \Gamma_1 \Gamma_2\right)}{\sqrt{\Gamma_1\Gamma_2\Gamma_3+ C^2 \left(\Gamma_2\Gamma_3 v_1^2+\Gamma_1\Gamma_3v_2^2+\Gamma_1\Gamma_2 v_3^2\right) }} \\
&+ p_T'
\text{diag}\left( \frac{\Gamma_2\Gamma_3}{\left[\Gamma_2-\Gamma_3 \right]^2} ,
\frac{\Gamma_1\Gamma_3}{\left[\Gamma_3-\Gamma_1 \right]^2},
\frac{\Gamma_1\Gamma_2}{\left[\Gamma_1-\Gamma_2 \right]^2}\right) .
\end{aligned}
\label{eq:BIX_geodesic_final}
\end{equation}
Together with the constraint $v_1^2+v_2^2+v_3^2=1$ this is all we need for numerical integration.
Note that all dependence on the Euler angles and their momenta has dropped out from  the equations of motion (\ref{eq:BIX_nondiag_eoms_rot_dust},\ref{eq:BIX_geodesic_final}).
The numerical method we use is described in Appendix~\ref{App:Numerical_analysis}.

\subsection{The tilted dust case}
\begin{figure}[!ht]
\centering
\includegraphics[width=7.0cm]{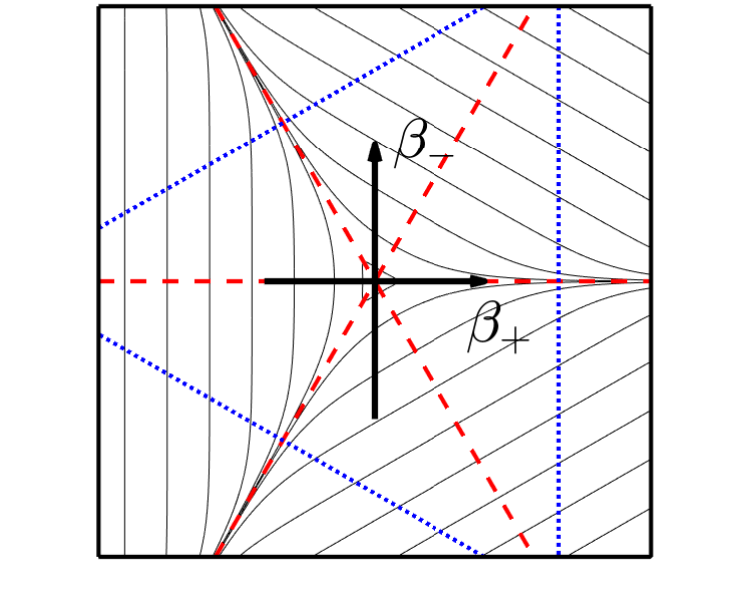}	
\caption{Potential picture for the tumbling case. The black contour lines represent the curvature potential while the dotted  (blue) and dashed (red) lines represent the rotation and centrifugal walls, respectively.}
\label{fig:BIX_potential_gen}
\end{figure}

A qualitative picture for the dynamics of the universe can be obtained by
considering the Hamiltonian
(\ref{eq:Hamiltonian+diffeo_constraints_dust}) in the quasi-Gaussian
gauge $N=\ee^{3\alpha}$, $N_i=0$.  The  Hamiltonian can then be
interpreted as the ``relativistic energy'' of a point particle
(called the universe point) with ``spacetime coordinates''
$(\alpha,\beta_+,\beta_-)$. The universe point is subject to the
forces generated by the dynamical potential
\begin{equation}
\frac{l_1^2}{I_1}+\frac{l_2^2}{I_2}+\frac{l_3^2}{I_3}
- \frac{\ee^{6\alpha}}{6} {}^{(3)}R
+ 2\ee^{3\alpha} p_T \sqrt{1+h^{ij}u_i u_j} \ ,
\end{equation}
which is depicted in Fig.~\ref{fig:BIX_potential_gen}.
 The contour lines of the curvature potential $-\frac{\ee^{6\alpha}}{12} {}^{(3)}R$ are represented by solid black lines. The curvature potential is  exponentially steep and takes its minimum at the origin $\beta_\pm=0$. When the universe evolves towards the singularity ($\alpha\rightarrow-\infty$), the  curvature potential walls move away from the origin while becoming {\it effectively} hard walls in the vicinity of the singularity.
The term $\ee^{3\alpha} p_T \sqrt{1+h^{ij}u_i u_j} $ can be
interpreted as three rotational potential walls. These potentials are
rather unimportant in the examination of the asymptotic dynamics,
since they move away from the origin with unit speed. The term
$\frac{l_1^2}{I_1}+\frac{l_2^2}{I_2}+\frac{l_3^2}{I_3}$ can be interpreted as three singular centrifugal potential walls.
They are represented by the dashed red lines.
Asymptotically close to the singularity, these walls are expected to become static.
In general, however, the centrifugal potential walls are dynamical and change in a complicated manner dictated by the geodesic equation (\ref{eq:BIX_geodesic_final}). The centrifugal walls will prevent the universe point from penetrating certain regions of the configuration space.
Misner \cite{Misner_1969b} and Ryan \cite{Ryan_1971a,Ryan_1971b}
employed these facts to obtain approximate solutions in a diagrammatic
form. The other Bianchi models can be treated in a similar way
\cite{Jantzen:2001me}.

\subsubsection{Special classes of solutions}

\begin{figure}[!ht]
\centering
\hspace{-2.5cm}
	\begin{subfigure}[t]{0.4\textwidth}
		\includegraphics[width=8cm]{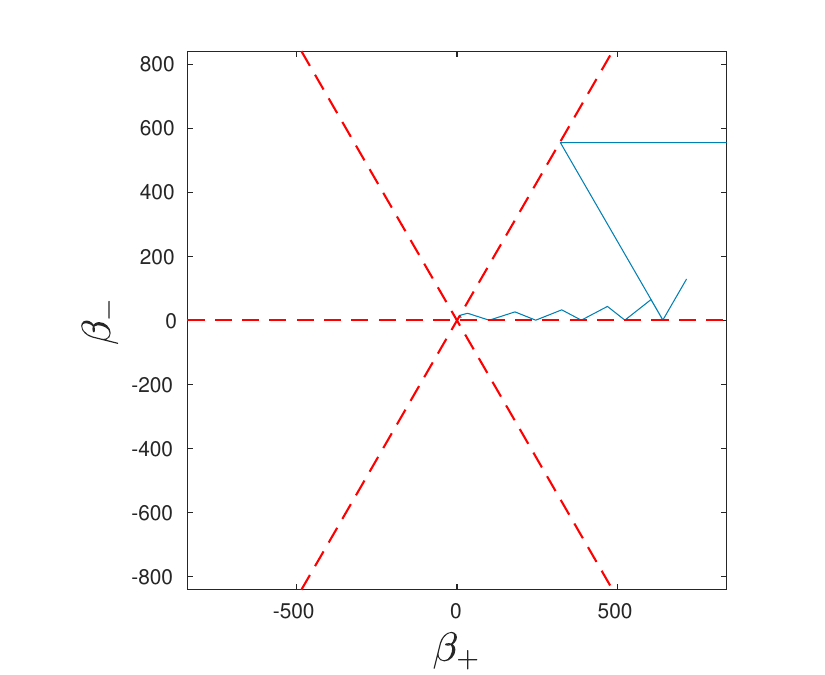}
	\end{subfigure}
	\qquad \quad
	\begin{subfigure}[t]{0.4\textwidth}
		\includegraphics[width=8cm]{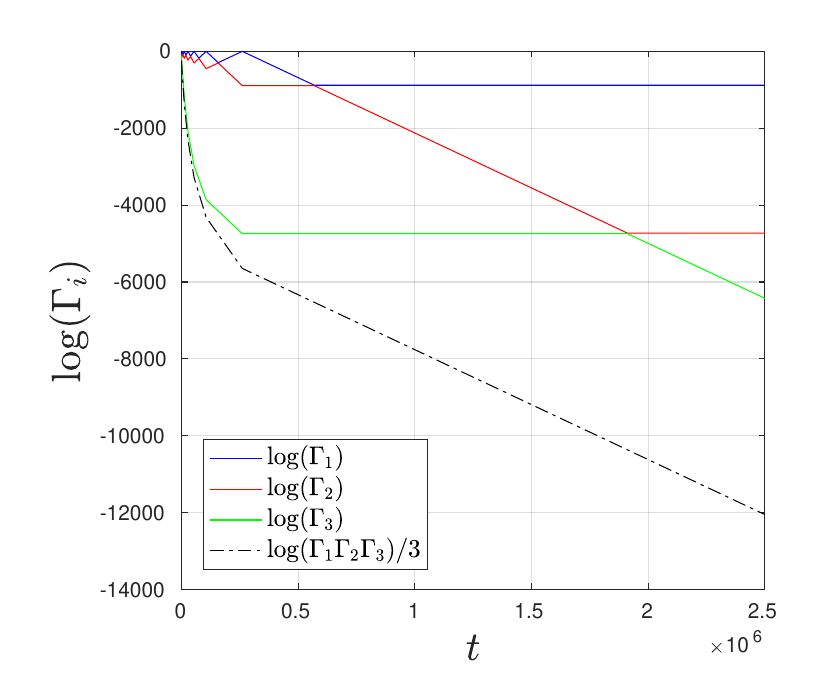}
	\end{subfigure}\\
	\caption{Numerical simulation of the tumbling case. In the
          region where $0<t<5\times 10^5$ (arbitrary units) one can
          see a typical Kasner era. The solution bounces in the valley
          formed between the curvature and one of the centrifugal
          potential walls. Increasing $t$ corresponds to evolution
          towards the singularity.} 
	\label{fig:BIX_tumbling2}
\end{figure}

Before doing the numerics, we comment on particular classes of solutions:
one class of solutions is obtained if we choose, for example, the
initial conditions $v_1=0=v_2$ and $v_3 = 1$. The geodesic equation
(\ref{eq:BIX_geodesic_final}) implies now that the velocities stay
constant in time. This implies that at all times $l_1= 0= l_2$ and
$l_3 = p_\psi = p_T C$.  This class of solutions is known as the
non-tumbling case.
Furthermore,  there are classes of solutions which are rotating
versions of the Taub solution.
These solutions should be divided into two subclasses: one class that
oscillates between the centrifugal walls and the curvature potential and one class that runs through the valley straight into the singularity.
We set
\begin{equation}
v_1=v_2=\frac{1}{2} \ , v_3=0 \quad \text{and} \quad \beta_- = 0  .
\end{equation}
For the $\Gamma_i$ variables it means that $\Gamma_1=\ee^{2\alpha}\ee^{2\beta_+}=\Gamma_2$ and $\Gamma_3=\ee^{2\alpha}\ee^{-4\beta_+}$.
With this choice we obtain $I_3=0$ and
$3I_1=3I_2=\sinh^2\left(3\beta_+ \right)$. Most importantly, the
geodesic equation (\ref{eq:BIX_geodesic_final}) is trivially
satisfied, that is, $v_1=v_2= 1/2$ and $v_3=0$ for all times.
When setting $C=0$ we obtain the diagonal case which contains the isotropic case of a closed Friedmann universe.
The simulation plotted in Fig.  \ref{fig:BIX_tumbling2} was performed
for the tumbling case, that is, the $v_i$ are chosen to be non-zero.

\subsubsection{The asymptotic regime close to the singularity}

\begin{figure}[!ht]
\centering
\hspace{-2cm}
	\begin{subfigure}[t]{0.4\textwidth}
		\includegraphics[width=8cm]{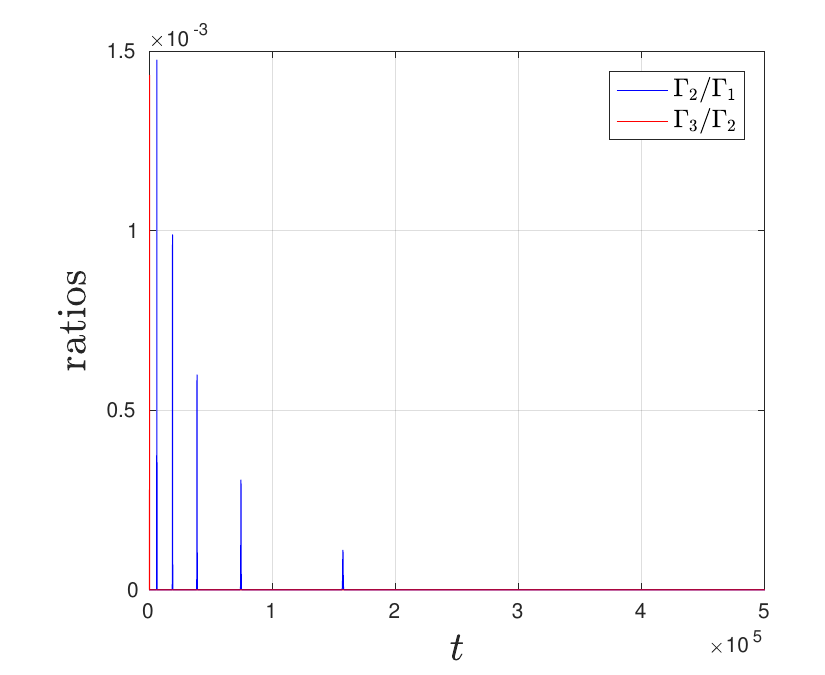}
	\end{subfigure}
	\qquad
	\begin{subfigure}[t]{0.4\textwidth}
		\includegraphics[width=8cm]{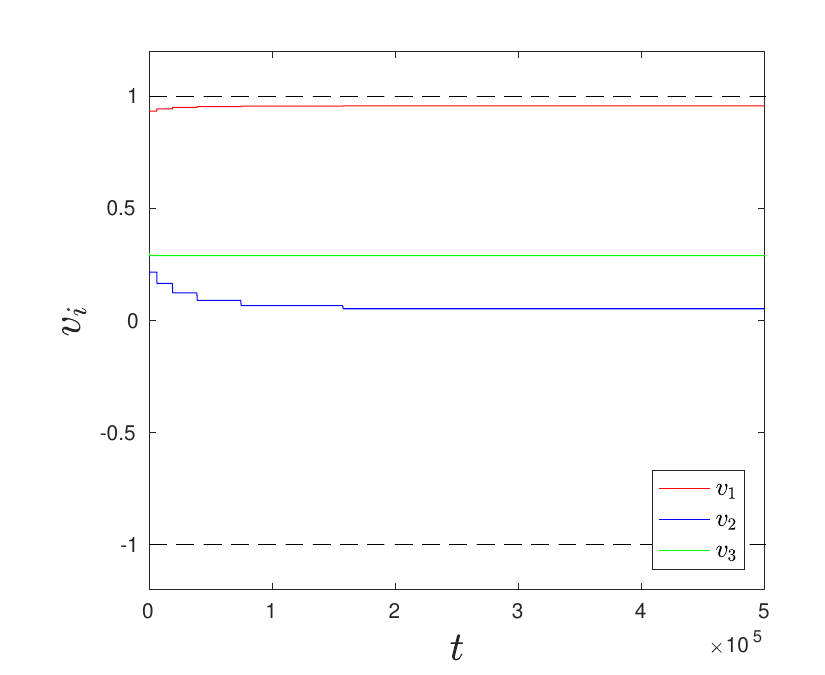}
	\end{subfigure}
	\caption{Plots of the ratios $\Gamma_2/\Gamma_1$,
          $\Gamma_3/\Gamma_2$ and the velocity $\vec{v}$. The plots
          correspond to the numerical solution presented in
          Fig. \ref{fig:BIX_tumbling2}. The peaks in the ratios appear
          during bounces of the universe point with the centrifugal
          walls.
As we can see, the height of these peaks decreases in the evolution
towards the singularity.}
	\label{fig:ratios_velocities}
\end{figure}

In order to simplify the dynamics of the general case, BKL made two
assumptions based on qualitative considerations of the equations of
motion.
The first assumption states that anisotropy of space grows without
bound. This means that the solution enters the regime
	\begin{equation}
		\Gamma_1\gg \Gamma_2\gg \Gamma_3 .
		\label{eq:BIX_Gamma_inequality}
	\end{equation}
The ordering of indices is irrelevant. In fact, there are six possible
orderings of indices which each correspond to the universe point being
constrained to one of the six regions bounded by the rotation and
centrifugal walls sketched in Fig.~\ref{fig:BIX_potential_gen}. The
region $\Gamma_1 > \Gamma_2> \Gamma_3$ corresponds to the right region
above the line $\beta_-=0$ in Fig.~\ref{fig:BIX_potential_gen}.
More precisely, the inequality (\ref{eq:BIX_Gamma_inequality}) means that
	\begin{equation}
	\Gamma_2 / \Gamma_1 \rightarrow 0 \quad \text{and}\quad
        \Gamma_3 / \Gamma_2 \rightarrow 0 \ .
	\end{equation}
Our numerical simulations support the validity of this assumption (see
the plot of the ratios $\Gamma_2 / \Gamma_1$ and $\Gamma_3 / \Gamma_2$
in Fig. \ref{fig:ratios_velocities}).
We provide plots of the two ratios $\Gamma_2/\Gamma_1$,
$\Gamma_3/\Gamma_2$ and the velocities $\vec{v}$ in order to provide a
sanity check of the approximation we will perform later on.

The second assumption made by BKL states that the Euler angles assume
constant values:
	\begin{equation}
	(\theta,\phi,\psi)\rightarrow (\theta_0,\phi_0,\psi_0) \ ,
	\end{equation}
that is, the rotation of the principal axes stops for all practical purposes and the metric becomes {\it effectively} diagonal.
The analysis of BKL
\cite{bkl} supports the consistency
of making both assumptions at the same time.
 {Similar heuristic considerations can possibly be applied to other
   Bianchi models as well \cite{Jantzen:2001me}}.
In the dust model under consideration, this assumption is equivalent
to the statement that the dust velocities $\vec{v}$ assume constant
values $\vec{v}\rightarrow \vec{v}^{(0)}$. Our numerical results
indicate that this is in fact true (see
Fig. \ref{fig:ratios_velocities}).
BKL then arrive at the simplified effective set of equations.

Let us now carry out the approximation and apply it to our equations of motion.   The kinetic term stays untouched during the approximation. The first step in the approximation is to ignore the rotational potential. In view of the strong inequality (\ref{eq:BIX_Gamma_inequality}), we approximate the curvature potential via
\begin{equation}
\Gamma_1^2+\Gamma_2^2+\Gamma_3^2-2(\Gamma_1\Gamma_2+\Gamma_3\Gamma_1+\Gamma_2\Gamma_3)
\approx \Gamma_1^2 \ .
\end{equation}
Furthermore, we approximate the centrifugal potential by
\begin{equation}
\frac{l_1^2}{I_1}+\frac{l_2^2}{I_2}+\frac{l_3^2}{I_3} \approx 12C^2 p_T^2 \left[
\frac{\Gamma_3}{\Gamma_2}\left( v^{(0)}_1 \right)^2+\frac{\Gamma_2}{\Gamma_1}\left( v^{(0)}_3 \right)^2\right] \ .
\end{equation}
Note that one centrifugal wall was ignored completely. Having Fig. \ref{fig:BIX_potential_gen} in mind, this approximation is well motivated since only two of the centrifugal walls are expected to have a significant influence on the dynamics of the universe point. After defining the new variables
\begin{equation}
a\equiv\Gamma_1\ , \quad
b\equiv2p_T'^2C^2\left( v^{(0)}_3\right)^2\Gamma_2  ,\quad
c\equiv4p_T'^4C^4\left( v^{(0)}_1 v^{(0)}_3\right)^2\Gamma_3 ,
\end{equation}
we arrive at a simplified Hamiltonian constraint and equations of motion,
\begin{equation}
\begin{aligned}
(\log a)^{\cdot}(\log b)^{\cdot }
+ (\log a)^{\cdot} (\log c)^{\cdot}
+ (\log b)^{\cdot} (\log c)^{\cdot} =
 a^2 + b/a +c/b   ,
\\
(\log a)^{\cdot \cdot}
= b /a - a^2   ,
\quad
(\log b)^{\cdot \cdot}
= a^2 - b/a + c/b   ,
\quad
(\log c)^{\cdot \cdot}  = a^2 - c/b  ,
\end{aligned}
\label{eq:BIX_nondiag_asympt}
\end{equation}
which coincides with the asymptotic form of equations obtained in \cite{bkl}.
Equations (\ref{eq:BIX_nondiag_asympt}) can now be treated by the numerical methods which we have used in the previous sections.
One must ensure that initial conditions are chosen such that
the simulation starts close to the asymptotic regime
(\ref{eq:BIX_Gamma_inequality}).

\section{Conclusions}

The numerical simulations indicate that the non-diagonal Bianchi IX  solutions, with tilted dust, evolve into the regime where
$\Gamma_1 \gg \Gamma_2 \gg \Gamma_3$ and $v_i\approx$const. The results motivate us to formulate the conjecture:

{\it Given a tumbling solution to the general Bianchi IX model filled with pressureless tilted matter, there exists
$t_0\in \mathbb{R}$ such that the solution is well approximated by a solution to the asymptotic equations of motion
for all times $t>t_0$ describing the vicinity of the singularity.}

To make the notion of ``approximation'' mathematically more precise, a suitable measure of the  ``distance'' on the
set of solutions is needed.
For this purpose, we propose to use the following simple measure:
\begin{equation}
\Delta (t)\equiv\sqrt{\left(\log \Gamma_1 (t) - \log \bar{a} (t) \right)^2
             +\left(\log \Gamma_2 (t) - \log \bar{b} (t)\right)^2
             +\left(\log \Gamma_3 (t) - \log \bar{c} (t) \right)^2} ,
\label{eq:error}
\end{equation}
where  $\{\Gamma_1,\Gamma_2,\Gamma_3\} $ denotes the numerical
solution to the exact equations of motion
(\ref{eq:BIX_nondiag_eoms_rot_dust})--(\ref{eq:BIX_geodesic_final}),
and \begin{equation}
a=\bar{a} \ ,  \quad
b=2p_T'^2C^2\left( v^{(0)}_3\right)^2\bar{b} \ ,\quad
c=4p_T'^4C^4\left( v^{(0)}_1 v^{(0)}_3\right)^2\bar{c} \
\end{equation}
denote the numerical solution to the asymptotic equations of motion  (\ref{eq:BIX_nondiag_asympt}).

We have evolved the exact system of equations from $t=0$ forward in
time  until $t=3 \times 10^6$. There we used the same initial
conditions as the ones we used to obtain the solution shown in Fig. \ref{fig:BIX_tumbling2}. We then took the final state at
$t=3 \times 10^6$ as an initial condition for the asymptotic system of equations and evolved it backwards in time towards
the re-bounce until $t=-980$.

\begin{figure}[!ht]
\centering
\hspace{-1.5cm}
	\begin{subfigure}[t]{0.45\textwidth}
		\includegraphics[width=8cm]{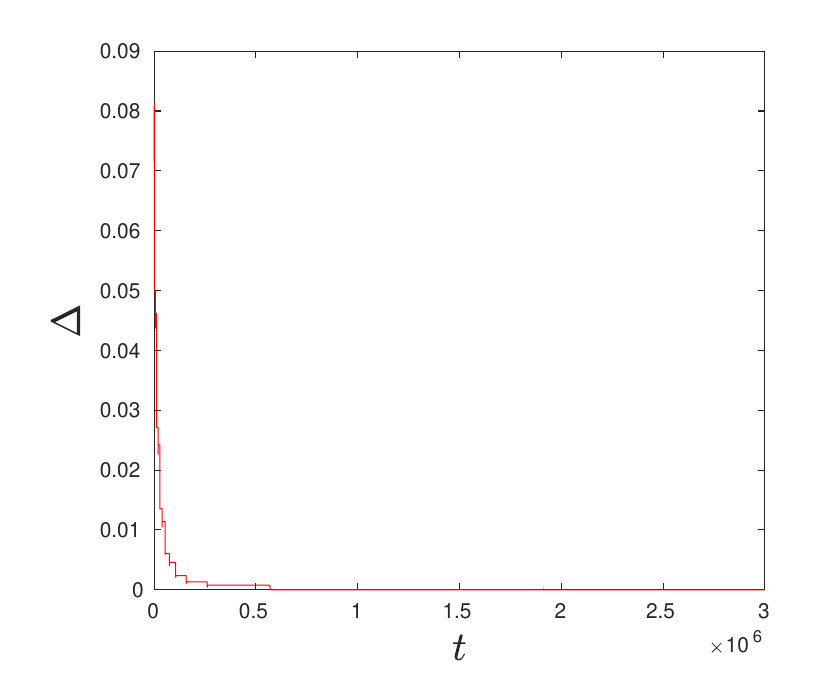}
		\subcaption{}
	\end{subfigure}
	\quad
		\begin{subfigure}[t]{0.45\textwidth}
		\includegraphics[width=8cm]{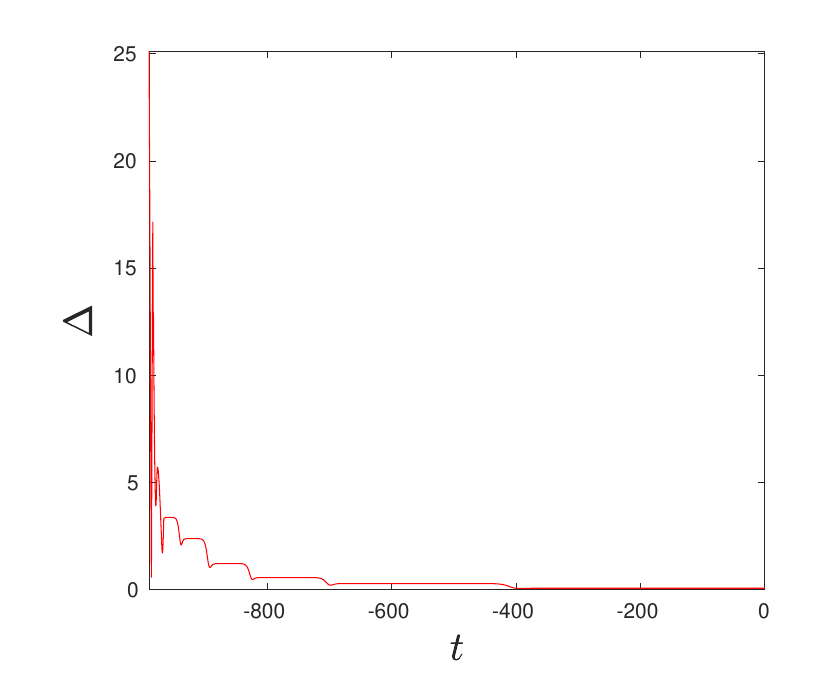}
		\subcaption{}
	\end{subfigure}
	\caption{The difference between the exact and the asymptotic solutions:
	(a) evolution towards the singularity, (b)  evolution away from the singularity.}
	\label{fig:BIX_Error}
\end{figure}

Fig. \ref{fig:BIX_Error} presents the measure (\ref{eq:error}) as a
function of time. We can see fast decrease of $\Delta$ with
increasing time (evolution towards the singularity) and fast increase
of $\Delta$ with decreasing time (evolution away from the
singularity).

Our numerical simulations give strong support to the conjecture
concerning the asymptotic dynamics  of the general Bianchi IX spacetime
put forward long ago by Belinski, Khalatnikov, and Ryan \cite{bkl}.
We remark that
approximating the diagonal or non-tumbling case by the asymptotic
dynamics (\ref{eq:BIX_nondiag_asympt}) is invalid (see \cite{ENW} for
more details).

It is sometimes stated
that ``matter does not
matter'' in the asymptotic regime, but this does not mean that
one is allowed to use the dynamics of the purely diagonal case.  One
only encounters an {\it
  effectively} diagonal case, which is expressed in terms of the
directional scale
factors $\{a,b,c\}$. So there exist serious differences between the purely
diagonal and effectively diagonal cases (see \cite{ENW} for more
details).

Employing the asymptotic form of the equations of motion may enable
one to study the chaotic behaviour and other properties of the
solution space for the general model.
 This is also important
for quantizing the general Bianchi IX model, where the quantization
of the exact dynamics seems to be quite difficult,
whereas the quantization of the asymptotic case seems to be feasible \cite{AGWP}.

\acknowledgments We are grateful to Vladimir Belinski and Claes Uggla
for helpful discussions. This work was supported by the 
German-Polish bilateral project DAAD and MNiSW, No. 57391638, 
``Model of   
stellar collapse towards a singularity and its quantization''.

\appendix

\section{Numerical analysis}
\label{App:Numerical_analysis}
Numerical simulations of the diagonal Bianchi IX model were already
carried out in the late 1980s and early 1990s (see
e.g. \cite{Berger_1990,Hobill_1991}; for a modern account, see \cite{Berger}).
Our main interest here is in the nondiagonal case.
With given initial conditions (respecting the constraints
$\mathcal{H}=0$), the system
(\ref{eq:BIX_nondiag_eoms_rot_dust}) and (\ref{eq:BIX_geodesic_final})
can  be integrated by using a suitable numerical method.  In this work
we employ the MATLAB R2016b solver ode113 \cite{matlab_ode}.
This code is an implementation of an Adams-Bashforth-Moulton method. It turns out to lead to the best results when compared to other MATLAB solvers. The relative error tolerance of the solver was chosen to be of the order $10^{-14}$.
We can integrate the equations of motion together with the geodesic
equation to obtain a numerical solution to the system. We set up
initial conditions at $t=0$  and evolve the system forward in time
towards the final singularity and away from the rebounce.

\begin{figure}[!ht]
\centering
\hspace{-1.5cm}
	\begin{subfigure}[t]{0.45\textwidth}
		\includegraphics[width=8cm]{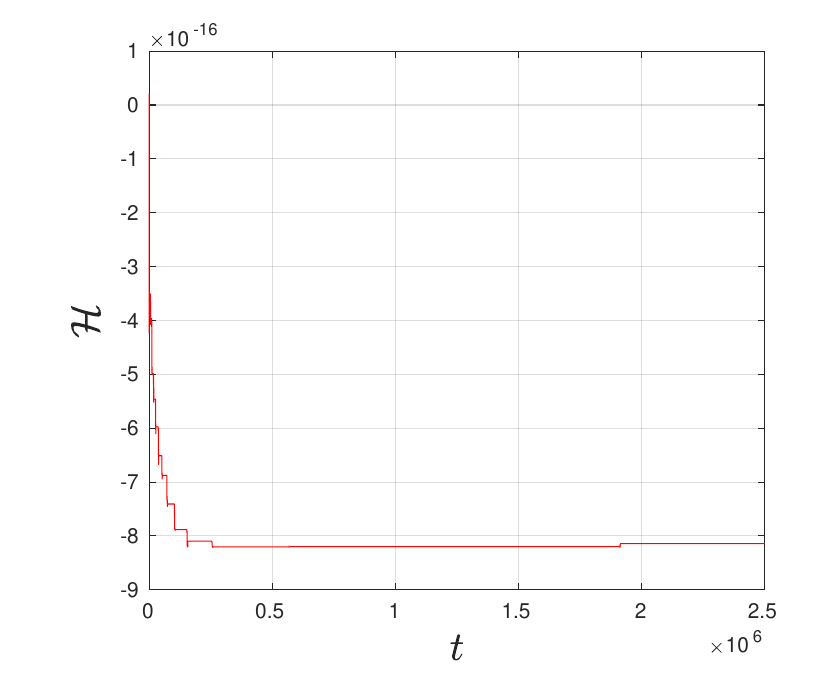}
		\subcaption{}
	\end{subfigure}
	\qquad
		\begin{subfigure}[t]{0.45\textwidth}
		\includegraphics[width=8cm]{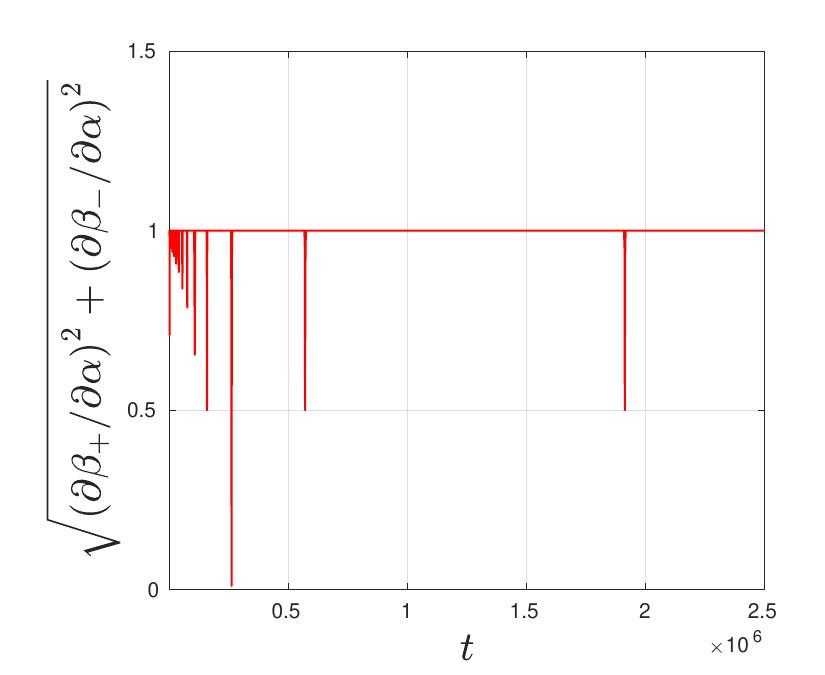}
		\subcaption{}
	\end{subfigure}
	\caption{Subfigure (a) shows the error in the Hamiltonian constraint. Subfigure (b) shows the speed of the universe point in the beta plane as measured in ``$\alpha$-time''. This plot can be viewed as another check of the numerics. Between two successive bounces from the potential walls this quantity should be close to one. If this ceased to be true it would indicate that the error in the Hamiltonian constraint becomes relevant and cannot be neglected in the approach towards the singularity (see \cite{Berger} for a more detailed discussion). Both plots correspond  to the numerical solution shown in Fig. \ref{fig:BIX_tumbling2}.}
	\label{fig:error_Hamiltonian_constraint}
\end{figure}

A major problem in numerical relativity is that  the Hamiltonian
constraint is not preserved exactly by the numerical procedure.
Similar to \cite{Berger_1990,Hobill_1991}, we find that the error in
the Hamiltonian constraint varies strongest after the start of the
simulation. Furthermore, it varies strongly when the evolution of
the universe approaches the point of maximal expansion.
While approaching the singularity, the error approaches an
approximately constant value. This can be seen in
Fig. \ref{fig:error_Hamiltonian_constraint}.
Therefore we can minimize the error when we choose the initial
conditions far away from the point of maximal expansion. Moreover, it
turned out that the error can be further reduced when constraining the
solver's maximally allowed time step size. This time step size should,
however, not be chosen too small since small time step sizes can drive
the
propagation of round off errors. Small step sizes are, of course, also numerically more expensive.
 By manually fine tuning the initial conditions and the maximally allowed time step size it was possible to keep the order of the error lower than $10^{-15}$.

Recall that the dynamics of Bianchi IX are chaotic,
that is, slightly changing initial conditions have a large effect on
the long time behaviour of solutions. Since the propagation of random
numerical errors cannot be avoided, we will be dealing with a
``butterfly effect'' and it should in general not be expected that our
numerical solution is an actual approximation of some exact solution
of the equations of motion when considering large time intervals.

\newcommand{\journal}{\rm}
\newcommand{\booktitle}{\em}
\newcommand{\vol}{\bf}
\newcommand{\papertitle}{\em}
\newcommand{\publisher}{\rm}

\end{document}